\documentstyle[12pt]{article}
\newcommand{\bb}{\begin{eqnarray}}
\newcommand{\ee}{\end{eqnarray}}
\begin{document}
\title{{Entropy of extremal black holes in asymptotically 
anti-de Sitter spacetime}}
\author{P. Mitra\thanks{e-mail mitra@tnp.saha.ernet.in}\\
Saha Institute of Nuclear Physics\\
Block AF, Bidhannagar\\
Calcutta 700 064, INDIA}
\date{hep-th/9807094}
\maketitle
\begin{abstract}
Unlike the extremal Reissner - Nordstr\"{o}m black hole in ordinary
spacetime, the one in anti-de Sitter spacetime is a minimum of action and
has zero entropy if quantization is carried out after extremalization.
However, if extremalization is carried out after quantization, then the
entropy is a quarter of the area as in the usual case.
\end{abstract}

\bigskip

While the entropy of ordinary ({\it non-extremal}) black holes has been
known to be given by a quarter of the horizon area for a long time, there
has been some uncertainty in the case of extremal black holes. The usual
derivations do not go through in a straightforward manner, and because of
the difference in topology of euclidean extremal black holes and euclidean
non-extremal black holes, one cannot fall back on extrapolation. In fact,
it has been suggested that extremal black holes should have zero entropy
\cite{HHR} even though the horizon area is nonzero.

On the other hand, some microscopic models have indicated that extremal
black holes could satisfy the area law just like non-extremal black holes.
One way out of this mismatch would be to say that the microscopic model is
wrong, but it is also possible to argue that the arguments of \cite{HHR} are
somewhat na\"{\i}ve.  Usually, when one quantizes a classical theory, one
tries to preserve the classical topology. In this spirit, \cite{HHR}  seeks
to have a quantum theory of extremal black holes based exclusively on
extremal topologies. As an alternative, one can try out a quantization where
a sum over topologies is carried out.  Then, in the consideration of the
functional integral, classical configurations corresponding to both
topologies must be included. The extremality condition can  subsequently be
imposed on the averages that result from the functional integration. It is
convenient, following \cite{GH} and \cite{york}, to use a grand canonical
ensemble. Here the temperature and the chemical potential for the charges
are supposed to be specified as inputs, and the average mass $M$ and charges
$Q$ of the black hole are outputs. So the actual definition of extremality
that is involved here for a Reissner - Nordstr\"{o}m black hole with one kind
of charge is $Q=M$.  This may be described  as {\it extremalization after
quantization}, as opposed to the usual approach of {\it quantization after
extremalization} \cite{GMthree}.  It was shown in \cite{GMthree} that
extremalization after quantization does lead to an entropy equal to a
quarter of the area.

Does the approach of quantization after extremalization in the case of the
Reissner - Nordstr\"{o}m black hole lead to zero entropy as suggested in
\cite{HHR}? Even that is not quite true \cite{GMcom}: the reason is that the
semiclassical approximation fails because the action does not have a stable
minimum there.

In view of recent interest in anti-de Sitter geometries, an investigation
has been made to determine whether anything more interesting happens if an
asymptotically anti-de Sitter version of the extremal Reissner -
Nordstr\"{o}m black hole is considered. It will be shown that a stable
minimum does occur in this case. Consequently, there is a sensible
semiclassical approximation, and as expected in \cite{HHR}, the entropy
vanishes if quantization is carried out after extremalization.  However, if
quantization is carried out first, the entropy is once again given by a
quarter of the area.

The   Reissner   -   Nordstr\"{o}m   black  hole   solution    of
Einstein's  equations  in free space with a negative cosmological
constant $\Lambda=-{3\over l^2}$ is given by (see {\it e.g.} \cite{louko})
\bb
ds^2=-hdt^2+h^{-1}dr^2+r^2d\Omega^2, ~A={Q\over r}dt,
\ee
with
\bb
h= 1-{r_+\over r}-{r_+^3\over l^2r}-{Q^2\over r_+r}
+{Q^2\over r^2}+{r^2\over l^2}.
\ee
The asymptotic form of this spacetime is anti-de Sitter.
There is an outer horizon  located at $r=r_+$.
The mass of the black hole is given by
\bb
M={1\over 2}(r_++{r_+^3\over l^2}+{Q^2\over r_+}).
\ee
It satisfies the laws of black hole thermodynamics with a temperature
\bb
T_H={1-{Q^2\over r_+^2} +{3r_+^2\over l^2}\over 4\pi r_+}\label{T}
\ee
and a potential
\bb
\phi={Q\over r_+}.
\ee
In general $r_+,Q$ are independent, but in the extremal case they get
related:
\bb
1-{Q^2\over r_+^2} +{3r_+^2\over l^2}=0.\label{ex}
\ee

The action for the euclidean version of the anti-de Sitter
Reissner - Nordstr\"{o}m
black hole on a four dimensional manifold ${\cal M}$ with a boundary
is given by
\bb
I&=&-{1\over  16\pi}\int_{\cal M} d^4x\sqrt g(R-2\Lambda)
+{1\over 8\pi}\int_{\partial
{\cal M}} d^3x\sqrt\gamma (K-K_0)\nonumber\\
&+&{1\over 16\pi}\int_{\cal M}
d^4x\sqrt g F_{\mu\nu} F^{\mu\nu}.\label{action}
\ee
Here  $\gamma$ is the induced metric on the boundary $\partial {\cal M}$
and $K$ the extrinsic curvature of the boundary. $K_0$ is to be
chosen to make the action finite.

The on-shell action for the black hole with the boundary taken at
$r=r_B$ and euclidean time integrated over from $0$ to $\beta$ is
\bb
&&{\beta\over 2l^2} (r_B^3-r_+^3) - {\beta\over 2}
\bigg(\sqrt{1-{r_+\over r}-{r_+^3\over l^2r}-{Q^2\over r_+r}
+{Q^2\over r^2}+{r^2\over l^2}} \nonumber\\ &&\times{d\over dr}\bigg[r^2
\sqrt{1-{r_+\over r}-{r_+^3\over l^2r}-{Q^2\over r_+r}
+{Q^2\over r^2}+{r^2\over l^2}}\bigg]\bigg)_{r=r_B}
+\nonumber\\
&&{\beta\over 2}r_B^2
\sqrt{1-{r_+\over r_B}-{r_+^3\over l^2r_B}-{Q^2\over r_+r_B}
+{Q^2\over r_B^2}+{r_B^2\over l^2}}(-K_0)\nonumber\\
&&-{\beta\over 2}Q^2(r_+^{-1}-r_B^{-1}).
\ee
To  keep this finite in the limit $r_B\to\infty$, it is necessary
to take
\bb
K_0=-{2\over l}-{l\over r_B^2}.
\ee
With this choice, the $r_B\to\infty$ limit of the action is
\bb
{\beta\over 2}(M-Q\phi-{r_+^3\over l^2}).
\ee
The corresponding entropy is calculated by equating $\beta$ times the
free energy with the action in the leading semiclassical approximation:
\bb
S=\beta(M-Q\phi)-I= {\beta\over 2}(M-Q\phi+{r_+^3\over l^2})=
{\beta r_+\over 4}(1-{Q^2\over r_+^2}+{3r_+^2\over l^2})
\ee
If $\beta$ is taken to be the reciprocal of (\ref{T}), this expression
simplifies to a quarter of the area:
\bb
S_{non-ex}=\pi r_+^2.
\ee
In the {\it extremal} case, where
there is no conical singularity in the euclidean metric,
the (vanishing) expression for the temperature is {\it not} used for $\beta$,
which is allowed to be finite. Then one gets the entropy to be
\bb
S_{ex}={\beta r_+\over 4}(1-{Q^2\over r_+^2}+{3r_+^2\over l^2})
=0\ee
because of (\ref{ex}).
All this is very similar to what happens in the ordinary Reissner -
Nordstr\"{o}m case, to which everything reduces in the limit $l\to\infty$.

Now we turn to a study of the action for off-shell configurations near the
black hole solution. For simplicity, only
a class of spherically symmetric
metrics \cite{york} is considered on ${\cal M}$:
\bb
ds^2=b^2d\tau^2+\alpha^2dr^2+r^2d\Omega^2,
\ee
with  the variable $r$ ranging between $r_+$ (the horizon) and $r_B$ (the
boundary), and $b, \alpha$ functions of $r$  only.  There  are
boundary conditions as usual \cite{york,GMthree,RNads}:
\bb
b(r_+)=0,~2\pi b(r_B)=\beta.
\ee
This corresponds to the convention of fixing the range of integration
of the euclidean time $\tau$ to be $2\pi$.
$\beta$  is the inverse temperature at the boundary of radius $r_B$.
There is another boundary condition involving $b'(r_+)$:
It reflects the extremal/non-extremal  nature  of  the  black  hole  and  is
therefore   different   for  the  two conditions:
\bb
{b'(r_+)\over\alpha(r_+)}&=&1{\rm ~in~non-extremal~case},\nonumber\\
&{\rm and}& 0 {\rm ~in~extremal~case}.
\ee

The  vector
potential is taken to be zero and the scalar potential  satisfies
the boundary conditions
\bb
A_\tau(r_+)=0, ~A_\tau(r_B)={\beta\phi\over 2\pi i}.
\ee

The action (\ref{action}) with  this  form  of  the  metric  depends  on the
functions $ b(r), \alpha(r)$ and $A_\tau(r)$:
\bb
I&=&{1\over      2}\int^{2\pi}_0      d\tau\int^{r_B}_{r_+}      dr
\bigg(-{2rb'\over\alpha}  -{b\over\alpha}  -\alpha  b   +\Lambda\alpha
br^2\bigg)         -{1\over         2}\int^{2\pi}_0              d\tau
\bigg[{(br^2)'\over\alpha}\bigg]_{r=r_+}\nonumber\\ &+& I_0
+ {1\over      2}\int^{2\pi}_0      d\tau\int^{r_B}_{r_+}      dr
{r^2\over\alpha b}{A'}_\tau^2.
\ee
$I_0$ is the contribution of the $K_0$ term in the action.
Variation of the functions $ b(r), \alpha(r)$ and $A_\tau(r)$ with proper
boundary  conditions  leads  to  reduced versions of
the  Einstein  -  Maxwell
equations. The solution of a subset of these equations,
namely the Gauss law and the Hamiltonian constraint, is given by
\cite{york,RNads}
\bb
{1\over\alpha}=\bigg(1-{r_+\over r}-{r_+^3\over l^2r}-{q^2\over r_+r}
+{q^2\over r^2}+{r^2\over l^2}\bigg)^{1/2},
\quad A'_\tau=-{iqb\alpha\over r^2},
\ee
with  $r_+$ and  $q$ arbitrary at this stage.
The reason why these parameters are not
expressed as functions of $\beta,\phi$ is that some of the equations
of motion and the corresponding boundary conditions
have not yet been imposed on the solution. Instead of that, the
action may be expressed in terms of $r_+,q$ and then
extremized with respect to $r_+,q$ as in \cite{york}.

The value of the action corresponding to the solution depends
on the boundary condition:
\bb
I= -\beta \bigg(r_B\sqrt{1-{r_+\over r_B}-{r_+^3\over l^2r_B}-
{q^2\over r_+r_B}
+{q^2\over r_B^2}+{r_B^2\over l^2}}&&+q\phi\bigg) +I_0-\pi r_+^2
\nonumber\\ &&{\rm ~for~non-extremal~bc},\nonumber\\
I= -\beta \bigg(r_B\sqrt{1-{r_+\over r_B}-{r_+^3\over l^2r_B}-
{q^2\over r_+r_B}
+{q^2\over r_B^2}+{r_B^2\over l^2}}&&+q\phi\bigg)+I_0
\nonumber\\ &&{\rm ~for~extremal~bc}.\label{I}
\ee
The first line is analogous to \cite{york,RNads},
where the non-extremal boundary condition was used
in connection with a semiclassically quantized  non-extremal  black
hole. The second line is similar to  the consequence of the
extremal boundary condition
used in connection with  a  semiclassically
quantized   extremal   black   hole
\cite{GMcom}. 

The above ``reduced action'' has to be extremized with respect to $q,r_+$
in order to impose the equations of motion ignored so far.
The form of $I_0$ is not important for this as it does not involve
$q,r_+$ when $r_B$ is large. The extremization with respect to $q$
yields the relation
\bb
{{q\over r_+}-{q\over r_B}\over\sqrt{
1-{r_+\over r_B}-{r_+^3\over l^2r_B}-{q^2\over r_+r_B}
+{q^2\over r_B^2}+{r_B^2\over l^2},
}}=\phi,
\ee
while extremization with respect to $r_+$ yields
\bb
{\beta (1-{q^2\over r_+^2}+{3r_+^2\over l^2})\over\sqrt{
1-{r_+\over r_B}-{r_+^3\over l^2r_B}-{q^2\over r_+r_B}
+{q^2\over r_B^2}+{r_B^2\over l^2},
}}&=& 4\pi r_+ {\rm ~for~non-extremal~bc},\nonumber\\
&{\rm but}& 0 {\rm ~for~extremal~bc}.
\ee
For non-extremal boundary conditions, these two relations can be used
to fix $q,r_+$ in terms of the specified values of  $\beta,\phi$;
they also show the expected forms of $\beta,\phi$ as functions of
$q,r_+$.
The nature of the extremum has been discussed in \cite{RNads}.
The entropy can be calculated by standard thermodynamical methods
and is found to be the expected $\pi r_+^2$ \cite{RNads}.

Much the same thing can be done for the extremal boundary condition, where,
however, the second equation is homogeneous and $\beta$ disappears from
the relations. This is not surprising: $q,r_+$ are not independent in
this case, but are related to each other by (\ref{ex}), and the temperature
is undetermined as there is no conical singularity \cite{HHR}.
The first relation can be written as
\bb
{1+{3r_+^2\over l^2}\over
1+{3r_+^2+2r_+r_B+r_B^2\over l^2}}=\phi^2.
\ee
This is reminiscent of the fact that $|\phi|$ has to be unity
in the usual extremal case. In the anti-de Sitter situation the
restriction on $|\phi|$ is only that it has to be less than unity: $r_+$
can then be sought to be determined in terms of $\phi$ 
by solving the quadratic:
\bb
r_+={\phi^2r_B\pm\sqrt{\phi^4r_B^2-3(1-\phi^2)(l^2 -\phi^2l^2-\phi^2r_B^2)}
\over 3(1-\phi^2)}.
\ee
There are values of $\phi$ for which this equation has only complex
solutions, and even when there are real solutions, one solution may be
negative.  A  positive  solution $r_+$  does not necessarily mean
that the extremum of the action is a minimum.

The matrix of second derivatives of the action with respect to
$q,r_+$ is real, symmetric and equal to
$\pmatrix{ {\beta\phi\over q}+{\beta\phi^3\xi\over q(1-\xi)} &
-{\beta\phi\over r_+\xi}\cr
-{\beta\phi\over r_+\xi}
& {\beta\phi({q^2\over r_+^2}+{3r_+^2\over l^2})\over q(1-\xi)}}$,
where $\xi\equiv\frac{r_+}{r_B}$. For the action to be a minimum
at the extremum, this matrix has to be positive definite, {\it i.e.,}
both  of  its  eigenvalues  have  to be positive. In view of  the
reality of the eigenvalues, this is equivalent to the requirement
that both the
trace and the determinant have to be positive. The trace is  seen
to  be  positive  if  $r_+<r_B$.  We  shall  consider 
this condition to be imposed. The determinant is, up to  a  positive
factor,
\bb
{3r_+^2\over l^2}- (1-\phi^2)\xi(1+{6r_+^2\over l^2}),
\ee
which  again can  be  made  positive by making $r_B$ large enough, {\it
i.e.,} $\xi$ small enough. $r_+$ is to be  held  fixed
if  an  adjustment of $r_B$  has to be made, 
which means  that the value of $\phi$ is such that
$r_+$ turns out to be  appropriately  small  in  comparison  with
$r_B$. Large solutions for $r_+$ also exist for appropriate $\phi$: 
they do not correspond to minima of the action.
Note that the above expression vanishes in the limit $l\to\infty$
if the corresponding condition $|\phi|=1$ is imposed,
thus confirming that the extremal Reissner - Nordstr\"{o}m black hole 
in asymptotically flat spacetime is not a minimum of the action.

The entropy corresponding to the saturation of the action by this
minimum is zero. This follows from the fact that \cite{HHR}
the action continues to be proportional to $\beta$ after the extremizing
values of $q,r_+$ are plugged in. Hence,
\bb
S=\beta^2{\partial (I/\beta)\over\partial\beta}=0.
\ee

The above statements refer to the quantized extremal black hole.
As indicated above, there is a possibility of quantizing the black
hole {\it before} extremizing it, {\it i.e.,} the
two topologies may be summed over in the functional integral \cite{GMthree}
and the extremality condition imposed afterwards on the averaged quantities.
The partition function is of the form
\bb
\sum_{\rm topologies}\int d\mu(r_+)\int d\mu(q) e^{-I(q,r_+)},
\ee
with   $I$   given    by    (\ref{I})    as    appropriate    for
non-extremal/extremal topology.

The  semiclassical approximation involves
replacing the double integral by the
maximum value of the integrand, {\it i.e.,} by the
exponential of the negative of the minimum  $I$.
We consider the variation of $I$ as $q,r_+$ vary in both topologies.
It  is  clear from (\ref{I})
that the non-extremal action can be made lower than the extremal one
because of the extra term $-\pi r_+^2$.
Consequently, the partition function is  to  be  approximated  by
$e^{-I_{min}}$,  where $I_{min}$ is the classical
action for the {\it non-extremal} case,
{\it minimized} with respect to $q,r_+$.
As in the non-extremal case, this leads to an entropy equal to a quarter
of the horizon area. Extremality is imposed eventually through the condition
(\ref{ex}) on $q,r_+$. 

Thus the entropy depends very  significantly
on   whether   quantization   is   carried   out   first  or
extremalization \cite{GMthree}: in the former case, the answer is 
a quarter of the area, and in the latter, zero.

\end{document}